\documentclass{cimento}
\usepackage{graphicx}
\usepackage{cite}

\title{Study of multi-neutron emission in the $\beta$-decay of $^{11}$Li}

\author{F.~Delaunay\from{LPC}\ETC,
N.~L.~Achouri\from{LPC},
A.~Algora\from{IFIC},
M.~Assi\'e\from{IPNO},
J.~Balibrea\from{CIEMAT},
K.~Banerjee\from{VECC},
C.~Bhattacharya\from{VECC},
M.~J.~G.~Borge\from{ISOLDE}\from{IEM},
D.~Cano-Ott\from{CIEMAT},
B.~Fern\'andez-Dom\'inguez\from{USC},
L.~M.~Fraile\from{Complutense},
J.~Gibelin\from{LPC},
M.~V.~Lund\from{Aarhus},
M.~Madurga\from{ISOLDE},
F.~M.~Marqu\'es\from{LPC},
I.~Marroquin\from{IEM},
T.~Mart\'inez\from{CIEMAT},
E.~Mendoza\from{CIEMAT},
N.~A.~Orr\from{LPC},
M.~P\^arlog\from{LPC},
X.~Pereira-L\'opez\from{LPC}\from{USC},
V.~Pestel\from{LPC},
K.~Riisager\from{Aarhus},
C.~Santos\from{CIEMAT},
M.~S\'enoville\from{SPhN},
O.~Tengblad\from{IEM}
\atque
V.~Vedia\from{Complutense}
}

\instlist{
\inst{LPC} LPC Caen, Normandie Universit\'e, ENSICAEN, UNICAEN, CNRS/IN2P3, Caen, France
\inst{IFIC} Instituto de F\'isica Corpuscular, CSIC-Universidad de Valencia, Valencia, Spain
\inst{IPNO} Institut de Physique Nucl\'eaire, CNRS/IN2P3, Universit\'e Paris-Sud, Orsay, France
\inst{CIEMAT} Centro de Investigaciones Energ\'eticas, Medioambientales y Tecnol\'ogicas, Madrid, Spain
\inst{VECC} Variable Energy Cyclotron Centre, Kolkata, India
\inst{ISOLDE} ISOLDE-PH, CERN, Geneva, Switzerland
\inst{IEM} Instituto de Estructura de la Materia, CSIC, Madrid, Spain
\inst{USC} Universidade de Santiago de Compostela, Santiago de Compostela, Spain
\inst{Complutense} Universidad Complutense, Grupo de F\'isica Nuclear, CEI Moncloa, Madrid, Spain
\inst{Aarhus} Department of Physics and Astronomy, Aarhus University, Aarhus, Denmark
\inst{SPhN} CEA Saclay, DRF/IRFU/DPhN, Gif-sur-Yvette, France
}

\begin{document}

\maketitle

\begin{abstract}
The kinematics of two-neutron emission following the $\beta$-decay of $^{11}$Li was investigated for the first time by detecting the two neutrons in coincidence and by measuring their angle and energy.
An array of liquid-scintillator neutron detectors was used to reject cosmic-ray and $\gamma$-ray backgrounds by pulse-shape discrimination.
Cross-talk events in which two detectors are fired by a single neutron were rejected using a filter tested on the $\beta$-1n emitter $^9$Li. A large cross-talk rejection rate is obtained ($> 95 \%$) over most of the energy range of interest.
Application to $^{11}$Li data leads to a significant number of events interpreted as $\beta$-2n decay. A discrete neutron line at $\approx$ 2 MeV indicates sequential two-neutron emission, possibly from the unbound state at 10.6 MeV excitation energy in $^{11}$Be.
\end{abstract}

\maketitle

\section{Introduction}

The $\beta$-decay of $^{11}$Li has attracted considerable interest for the last decades, owing to the halo nature of $^{11}$Li and to its various delayed particle emission modes mainly due to the high $Q_\beta$-value (20.6 MeV) together with the weakly bound character of the $^{11}$Be daughter ($S_n=0.50$ MeV).
In addition to delayed $\gamma$-rays \cite{BorgeBetaGamma,FynboBetaGamma,SarazinBetaGamma,MattoonBetaGamma}, emissions of one \cite{Aoi,Morrissey,Hirayama}, two \cite{Azuma2n} and three neutrons \cite{Azuma3n}, and of charged particles \cite{RaabeBetaDeuteron,MadurgaBetaCPNPA,MadurgaBetaCPPLB} are known to occur.

Very little is known about $\beta$-2n emission apart from the emission probability, $P_{2n} = 4.2(4) \%$ \cite{BorgeBetaGamma}, which is one of the largest known values.
In particular, no direct detection of the two neutrons in coincidence, with measurement of their energy and angle, has been undertaken.
Tentative $\beta$-2n paths have been proposed from single-neutron transitions \cite{Hirayama}. However, these proposed paths imply a $P_{2n}$ emission probability larger than the accepted value of $4.2(4) \%$.
Similarly to two-proton decay, two-neutron emission could be direct or proceed by sequential emission through intermediate states. In the case of a direct decay, possible correlations between the two neutrons would be of considerable interest,
in particular as a probe of the configuration of the neutrons in the decaying state.

When directly detecting two delayed neutrons in coincidence, one difficulty arises from random coincidences involving backgrounds of $\gamma$-rays and cosmic muons. Another difficulty is due to cross-talk events in which a single neutron fires a first detector, to be then scattered to another detector and detected.
To overcome these difficulties, we have used a modular array of liquid scintillator neutron detectors.
The liquid scintillator allows the rejection of $\gamma$-ray and muon backgrounds through pulse-shape discrimination (PSD), while the modular nature of the array allows for the application of offline cross-talk filters \cite{MarquesCT}.
In addition, the neutron angle measurement is provided by the granularity of the array and the neutron energy is measured via the time-of-flight (TOF).

\section{Experiment}

The experiment was performed at the CERN-ISOLDE facility. The $^{11}$Li nuclei were produced by bombarding a Ta-foil target with 1.4 GeV protons. The products were ionized by a Re surface ionization source together with the RILIS laser ionization system, extracted at an energy of 30 keV, mass separated
and guided to a 0.4-mm thick Al foil where they were implanted.
The beam had an intensity of $\approx$ 1300 ions/s.
The $\beta$-rays were detected by a thin cylindrical plastic scintillator surrounding the implantation foil.
The detection of a $\beta$-ray gave the ``start'' signal for the measurement of the neutron time-of-flight, while the ``stop'' was given by the neutron detection.
The neutrons were detected in 39 modules, each based on a 5-cm thick cell of NE213-BC501A liquid scintillator \cite{Martinez,EDEN},
arranged in two arrays.
A ``near array'' comprising 29 modules at 1.5 m from the implantation point provided for the detection of neutron-neutron coincidences with a moderate energy resolution (7\%), whilst a ``far array'' consisting of 10 modules at 2.5 m provided for the detection of single neutrons with a better resolution (4\%) at the expense of a reduced solid angle coverage.
All detectors were coupled to the FASTER digital electronics and data acquisition system \cite{FASTER} developed at LPC Caen.

\section{$\beta$-delayed two-neutron emission from $^{11}$Li}

Cross-talk events were rejected using the ``strong'' filter of \cite{MarquesCT}, which maximises the rejection rate by assuming electron recoils.
The filter was tested on data from the $\beta$-1n emitter $^9$Li, for which events with two neutron hits were considered as cross-talk.
To insure good PSD, an offline threshold of 75 keVee was used, which corresponds to a neutron energy of 530 keV.
The cross-talk probability was computed as $P_{CT} = N_{CT}/N_n$, with $N_{CT}$ the number of cross-talk events given by the number of events with two neutron hits, and $N_n$ the number of neutron singles. The filter rejection efficiency was calculated as $R_{CT} = 1 - N_{CT,unid}/N_{CT}$, with $N_{CT,unid}$ the number of cross-talk events unrejected by the filter.
The cross-talk probability $P_{CT}$ is found to increase with the neutron energy $E_n$, consistently with our dedicated cross-talk measurements \cite{SenovillePhD}, from 0.027(3)\% for the $E_n \in [1.1, 1.8 \mbox{ MeV}]$ to 0.30(3)\% for $E_n \in [4.5, 10.1 \mbox{ MeV}]$.
The filter efficiency $R_{CT}$ is higher than 95\%
and largely independent of $E_n$ over the range 1.8-10.1 MeV.
It is slightly lower (92(3)\%) for $E_n \in [1.1, 1.8 \mbox{ MeV}]$
where $P_{CT}$ is the smallest.

\begin{figure}[htbp]
\begin{center}
\includegraphics[width=6.6cm, trim = 0.8mm 0mm 9mm 0mm, clip=true]{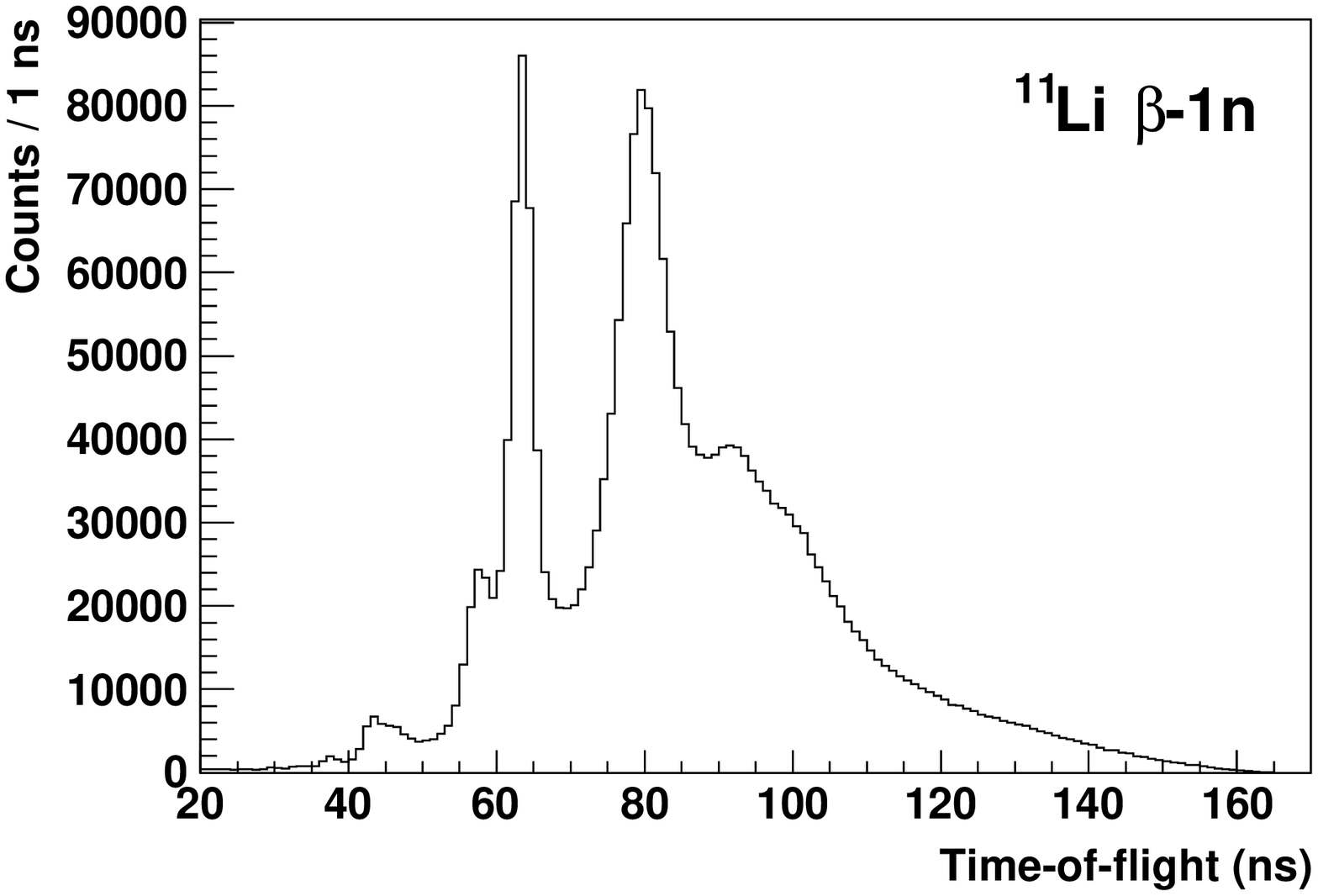}
\includegraphics[width=6.6cm, trim = 0.0mm 0mm 9.8mm 0mm, clip=true]{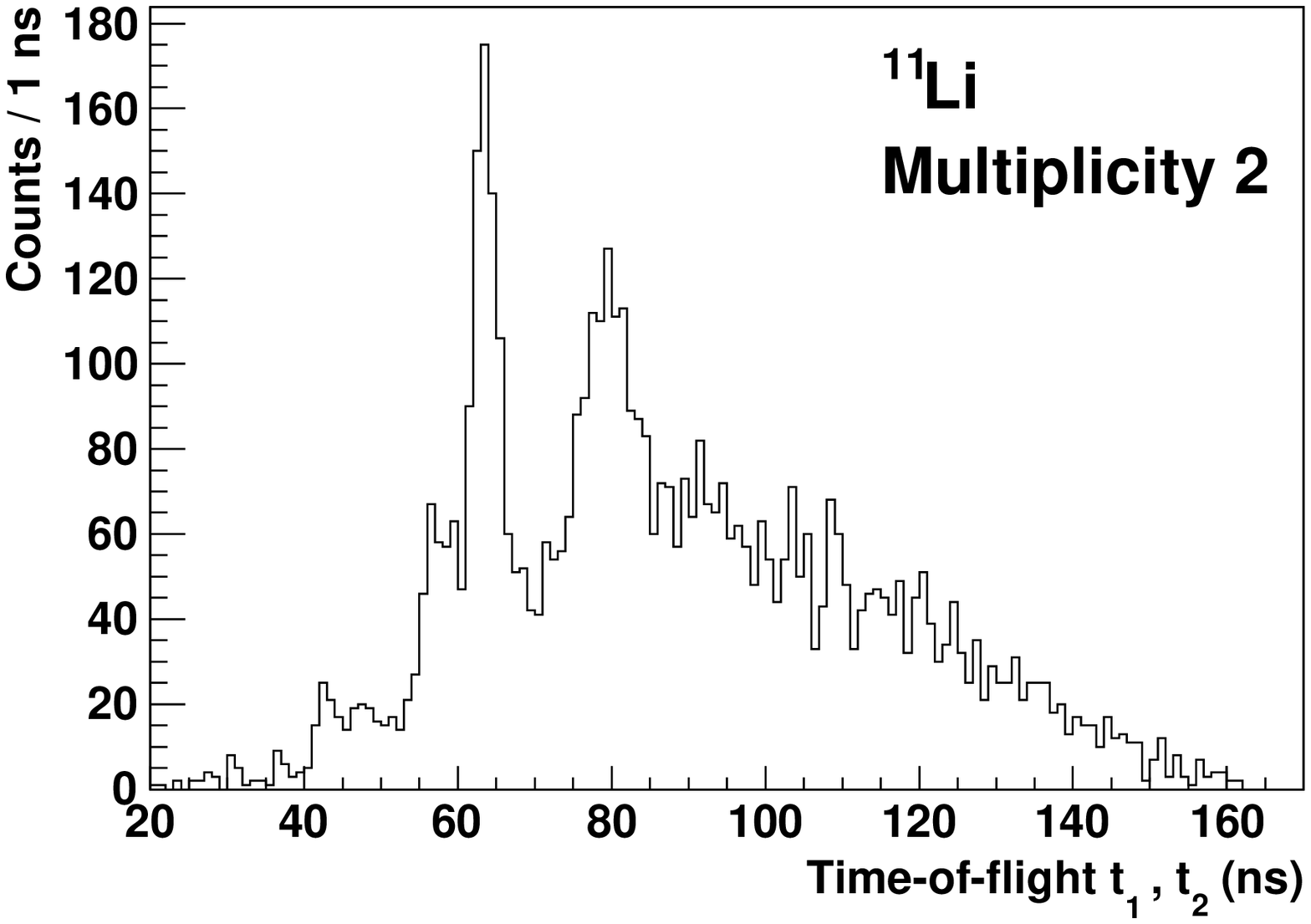}
\includegraphics[width=6.6cm, trim = 0.8mm 0mm 9mm 0mm, clip=true]{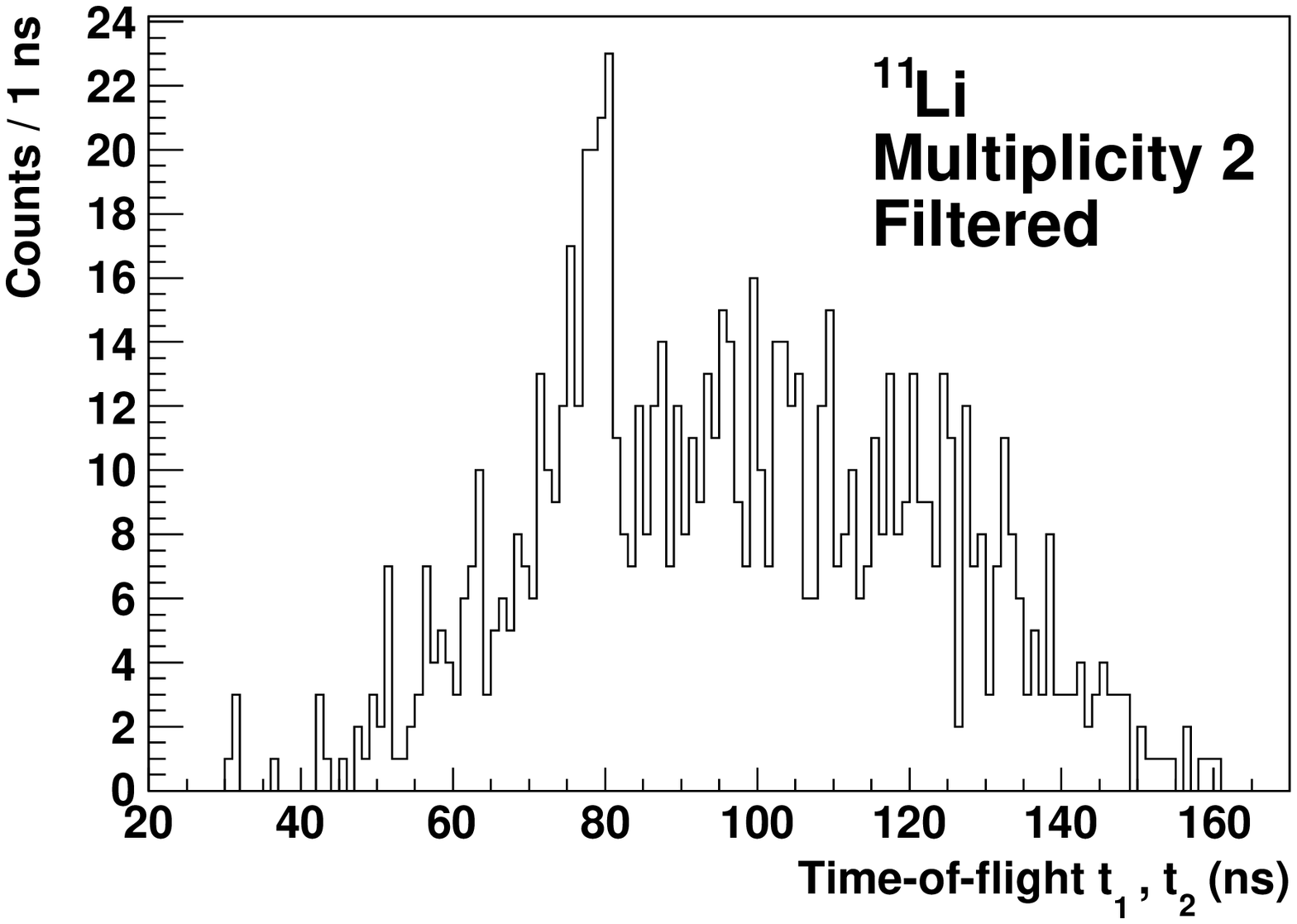}
\includegraphics[width=6.6cm, trim = 0.0mm 0mm 9.8mm 0mm, clip=true]{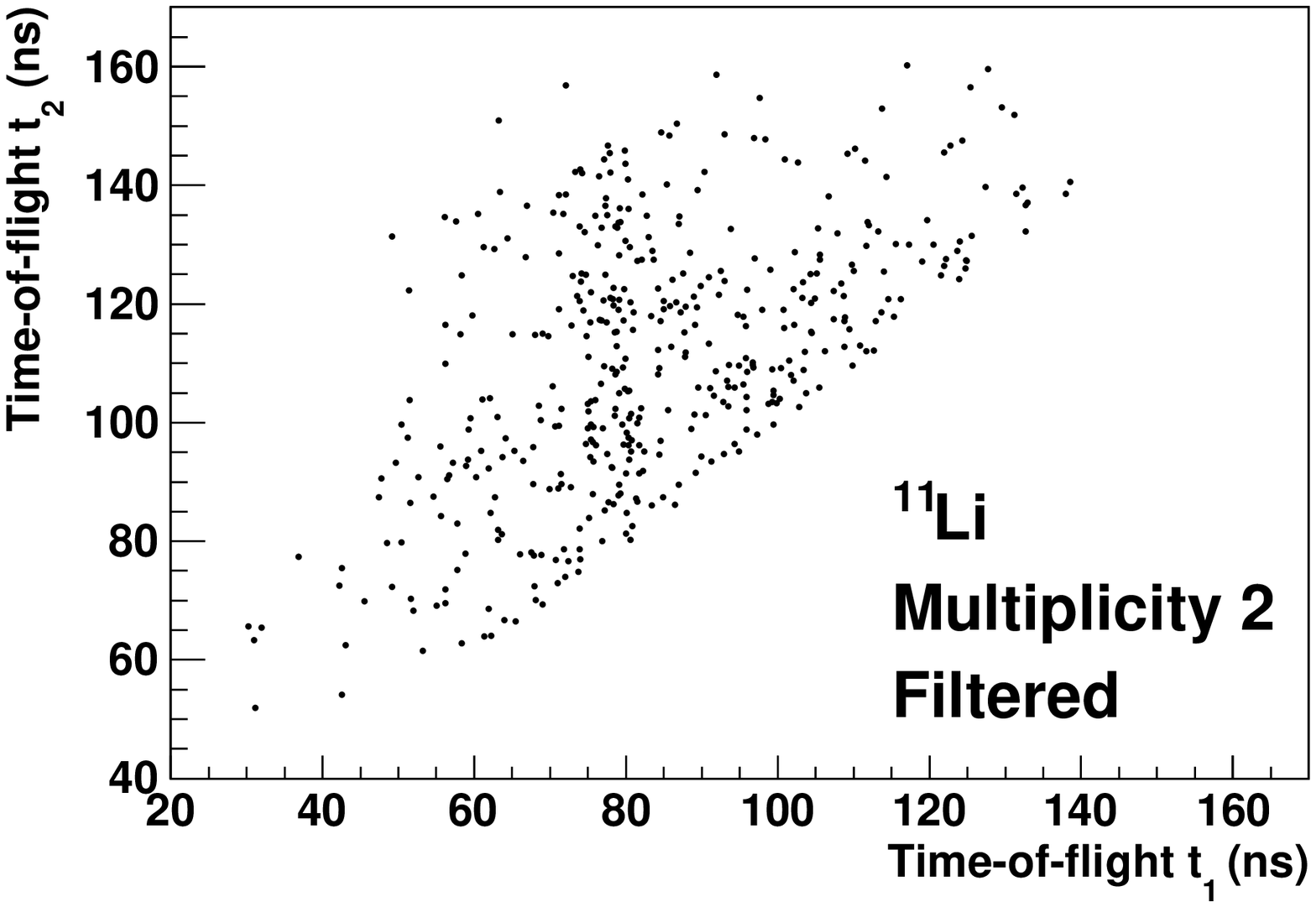}
\caption{Time-of-flight spectra of $^{11}$Li $\beta$-delayed neutrons detected in the near array. Top left: single neutron events. Top right: Multiplicity-2 events before cross-talk rejection. Bottom left: Multiplicity-2 events after cross-talk rejection. Bottom right: Time-of-flight of the second hit $t_2$ \textit{vs.} time-of-flight of the first hit $t_1$, after cross-talk rejection.}
\label{Fig11Li}
\end{center}
\end{figure}

The TOF spectrum of single neutrons following $^{11}$Li $\beta$-decay is shown on the upper left panel of fig. \ref{Fig11Li}, while
the upper right panel shows the TOF spectrum of events with two neutron hits. In the latter, peaks at $\approx$ 42, 58, 63 and 79 ns correspond to known single-neutron lines \cite{Aoi,Morrissey,Hirayama} visible on the singles spectrum and as such are due to cross-talk.

The lower left panel of fig. \ref{Fig11Li} presents the TOF spectrum of multiplicity-2 events after cross-talk rejection.
The number of these events is larger than the expected number of unrejected cross-talk events computed from the number of $^{11}$Li $\beta$-1n events using $P_{CT}$ and $R_{CT}$ measured with $^9$Li, and the difference is statistically significant, from $\approx 2 \sigma$ for $t_1 \in [35, 61 \mbox{ ns}]$ to $>5 \sigma$ for $t_1>68$ ns, $t_1$ being the TOF of the first neutron.
Therefore, these events can be interpreted as two-neutron emission following $^{11}$Li $\beta$-decay.
In particular, a peak at 79 ns is visible in the $\beta$-2n spectrum. As pointed out above, the number of 2n events is significantly larger than the number of unrejected cross-talk events. Furthermore, no clear peak is visible at 63 ns despite a larger cross-talk probability and a similar filter rejection efficiency compared to 79 ns. For these reasons the peak at 79 ns in the $\beta$-2n spectrum can not be a cross-talk remnant and it should then correspond to a neutron line from sequential two-neutron emission.

The lower right panel of fig. \ref{Fig11Li} shows the TOF of the slower neutron $t_2$ as a function of the TOF of the faster neutron $t_1$.
It shows that the line at 79 ns corresponds to the faster neutron, and that no narrow structure is seen in $t_2$ in coincidence with this line.
The TOF of 79 ns gives a neutron energy of 2 MeV. This line could correspond to either
of the ${^{11}\mbox{Be}} \rightarrow {^{10}\mbox{Be}}+n$ and ${^{10}\mbox{Be}} \rightarrow {^9\mbox{Be}}+n$ transitions of the sequential decay.
The reconstructed $^{11}$Be excitation energy for the events with $70 < t_1 < 85$ ns is concentrated in the range 9.8-11.8 MeV with a centroid at $\approx$ 10.7 MeV.
The emitting state might thus be the unbound state at 10.6 MeV excitation energy in $^{11}$Be.
Apart from the 79-ns line, the $t_2$ \textit{vs.} $t_1$ spectrum does not show any narrow structure. We can not at present conclude whether this continuous spectrum is due to unresolved neutron transitions from sequential emission or to the presence of direct emission, or to a combination of both.

To emphasise the crucial role of neutron-$\gamma$ discrimination in the selection of the $\beta$-2n events, we note that the same analysis without discrimination leads to a rejection efficiency of the cross-talk filter smaller than 25\%, and to a number of filtered multiplicity-2 events no longer significantly larger than the number of unrejected cross-talk events.

\section{Conclusions}

We have investigated for the first time the kinematics of two-neutron emission following $^{11}$Li $\beta$-decay by detecting the two neutrons in coincidence and measuring their energy and angle with a time-of-flight array.
The use of liquid scintillator detectors allowed rejection of random coincidences involving $\gamma$-rays and cosmic muons.
With data from the $\beta$-1n emitter $^9$Li, the rejection efficiency of the cross-talk filter was determined to be larger than 95\% over most of the neutron energy range of interest ($\approx$ 2-10 MeV).
A number of two-neutron events significantly larger than the number of unrejected cross-talk events was observed and these events were thus interpreted as $^{11}$Li $\beta$-2n emission. A 2-MeV neutron line indicates sequential two-neutron decay, possibly from the state at 10.6 MeV excitation energy in $^{11}$Be. The rest of the two-neutron time-of-flight spectrum is structureless and could correspond to unresolved neutron lines from sequential decay or to direct decay, or to a combination of both.

\end{document}